\documentclass[10pt,aps,prl,preprintnumbers,superscriptaddress,showpacs,twocolumn,floatfix]{revtex4-1}
\usepackage{graphicx}

\newcommand{\bfm}[1]{\mbox{\boldmath$#1$}}

\newcommand{\gsim}{\;\rlap{\lower 3.5 pt \hbox{$\mathchar \sim$}} \raise 1pt \hbox {$>$}\;}
\newcommand{\lsim}{\;\rlap{\lower 3.5 pt \hbox{$\mathchar \sim$}} \raise 1pt \hbox {$<$}\;}

\begin{document}

\title{
\boldmath Bottomonium Hyperfine Splitting
on the Lattice and in the Continuum.
\unboldmath}
\author{M. Baker}
\affiliation{Department of Physics, University of Alberta, Edmonton, Alberta T6G 2J1, Canada}
\author{A.A. Penin}
\affiliation{Department of Physics, University of Alberta, Edmonton, Alberta T6G 2J1, Canada}
\affiliation{Institut f\"ur Theoretische Teilchenphysik,
Karlsruhe Institute of Technology, 76128 Karlsruhe, Germany}
\author{D. Seidel}
\affiliation{Theoretische Physik 1, Universit\"at Siegen, 57068 Siegen, Germany}
\author{N. Zerf}
\affiliation{Department of Physics, University of Alberta, Edmonton, Alberta T6G 2J1, Canada}

\preprint{ALBERTA-THY-03-15, QFET-2015-16, SI-HEP-2015-13}

\begin{abstract}
We revise the  analysis of the bottomonium hyperfine splitting within the
lattice nonrelativistic  QCD. The Wilson coefficients of the radiatively
improved lattice  action are evaluated by a semianalytic approach based on the
asymptotic expansion about the continuum limit.  The nonrelativistic
renormalization group is used to estimate  the high-order radiative corrections.
Our result for the $1S$ hyperfine splitting is
$M_{\Upsilon(1S)}-M_{\eta_b(1S)}=52.9\pm 5.5~{\rm MeV}$. It  reconciles the
predictions of the continuum and lattice QCD and is in  very good agreement
with the most accurate experimental measurement by the Belle collaboration.
\end{abstract}
\maketitle
The bottomonium hyperfine splitting  defined by the mass difference $E_{\rm
hfs}=M_{\Upsilon(1S)}-M_{\eta_b(1S)}$ has been a subject of much controversy
since the first observation  of the spin-singlet $\eta_b$ state in radiative
decays of the $\Upsilon(3S)$ mesons by the Babar collaboration
\cite{Aubert:2008ba}. The measured value of the hyperfine splitting overshot the
predictions of  perturbative QCD \cite{Kniehl:2003ap} by almost  a factor of
two,  well  beyond the experimental and theoretical uncertainty bands, see
Table~\ref{tab::tab1}. Further experimental  studies
\cite{Aubert:2009as,Dobbs:2012zn,Bonvicini:2009hs} were consistent with the
initial measurement,  while the Belle collaboration  reported  a significantly
lower value of the splitting  with higher  experimental precision
\cite{Mizuk:2012pb}. On the  theory side the most accurate estimates of the
hyperfine splitting are obtained  from  lattice simulations within the
effective theory of nonrelativistic QCD (NRQCD).  This method is entirely based
on first principles, allows for  simultaneous treatment of dynamical heavy
and light quarks and gives a systematic account of the long-distance
nonperturbative effects of the strong interaction. The first analysis
\cite{Dowdall:2011wh} with fully incorporated one-loop  radiative corrections
\cite{Hammant:2011bt}  favored the larger value of the splitting
\cite{Aubert:2008ba}. The most recent analysis \cite{Dowdall:2013jqa} includes
the leading relativistic corrections and gives a lower value, which is close to
the  PDG average \cite{Beringer:1900zz} but nevertheless not consistent with
Ref.~\cite{Kniehl:2003ap}. This might indicate a serious failure of perturbative
QCD in the description of the bottomonium ground state in clear conflict with the
general concept of the  heavy quarkonium dynamics. Thus the current experimental
and theoretical status of the bottomonium hyperfine splitting remains ambiguous
and sets up  one of the most interesting open problems in the QCD  theory of
hadrons, which yet inspired a discussion about possible signal of physics
beyond the standard model  \cite{Domingo:2009tb}.

In this article we revise the analysis of the radiative corrections to the
lattice NRQCD action. We develop a semianalytical approach based on the
asymptotic expansion about the continuum limit \cite{Becher:2002if}, which
provides a very powerful tool for the radiative  improvement of lattice
NRQCD. Our result for the one-loop Wilson coefficient of the effective
spin-dependent four-quark interaction significantly differs from the result of
the previous calculation \cite{Hammant:2011bt}  used in the subsequent analyses
\cite{Dowdall:2011wh,Dowdall:2013jqa}, which leads to a sizable reduction of the
lattice NRQCD prediction for the hyperfine splitting. We  give an estimate of
the higher order radiative corrections by evaluating the two-loop
double-logarithmic terms  within  the nonrelativistic renormalization group
approach \cite{Penin:2004ay,Penin:2004xi}. The main result of this paper is a
new theoretical  value  for bottomonium hyperfine splitting,
Eq.~(\ref{eq::fin}).

The idea of the NRQCD approach \cite{Caswell:1985ui,Bodwin:1994jh} is to
separate the hard modes, which require a fully relativistic analysis, from
the nonrelativistic soft modes. The dynamics of the soft modes is governed by
the effective nonrelativistic action given by a series in heavy quark
velocity $v$, while the contribution of the hard modes is encoded in the
corresponding Wilson coefficients. The  nonrelativistic action can be applied in
a systematic perturbative analysis of the heavy quarkonium spectrum
\cite{Brambilla:1999xj,Kniehl:2002br,Penin:2002zv}. At the same time the action
may be used for lattice simulations of the heavy quarkonium states
\cite{Thacker:1990bm,Lepage:1992tx}. The latter approach gives  full control
over nonperturbative long-distance effects and can be used  for the description
of excited states where perturbative QCD is not applicable.

The hyperfine splitting {\it i.e.} the splitting between  the spin-singlet and
spin-triplet states is generated by the spin-dependent part of the NRQCD
Lagrangian. To order ${\cal O}(v^4)$ it reads (see {\it e.g.}
\cite{Pineda:1998kj,Pineda:1998kn})
\begin{equation}
{\cal L}_{\sigma}={c_F\over 2m_q}\psi^\dagger
{\bfm B}{\bfm \sigma}\psi +(\psi\to\chi_c)
+d_\sigma {C_F\alpha_s\over m_q^2}\psi^\dagger {\bfm \sigma}\psi \chi_c^\dagger
{\bfm \sigma}\chi_c,
\end{equation}
where $\bfm B$ is the chromomagnetic field,  $m_q$  and $\alpha_s$
are the  heavy quark mass and the strong coupling
constant, the  $SU(N_c)$ color group factor is $C_F=(N_c^2-1)/(2N_c)$, $\psi$
($\chi_c$) are the nonrelativistic Pauli spinors of quark (antiquark)   field,
and we have projected the four-quark interaction on the color-singlet state.
The Wilson coefficients $c_F$ and $d_\sigma$  logarithmically depend on
the factorization scale $\mu_f$ which separates the hard and the soft momentum
contributions.  This dependence can be predicted to all orders of perturbation
theory by renormalization group methods. In lattice NRQCD the natural
factorization scale is given by the inverse lattice spacing $a$. The radiative
improvement of the action is therefore mandatory for the correct continuum limit.

\begin{figure}[t]
\begin{tabular}{ccc}
\includegraphics[width=2.4cm]{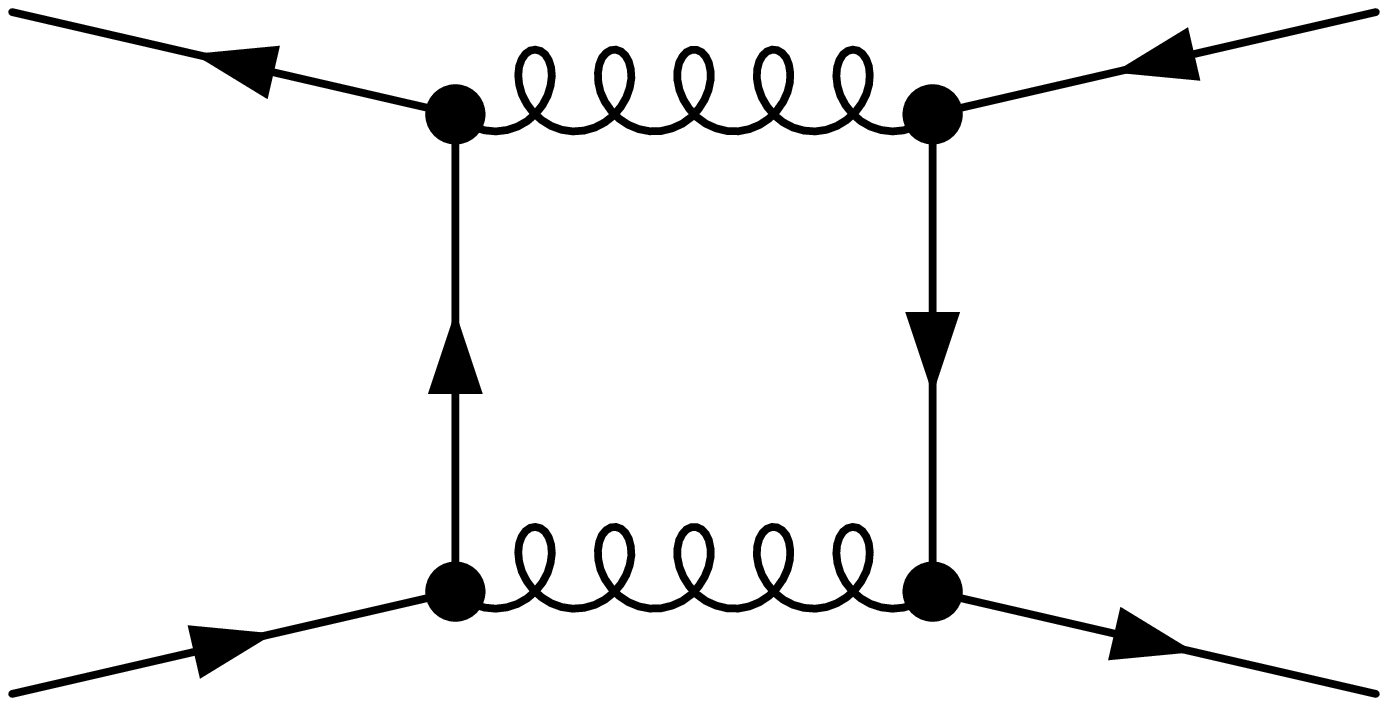}&
\hspace*{2mm}\includegraphics[width=2.4cm]{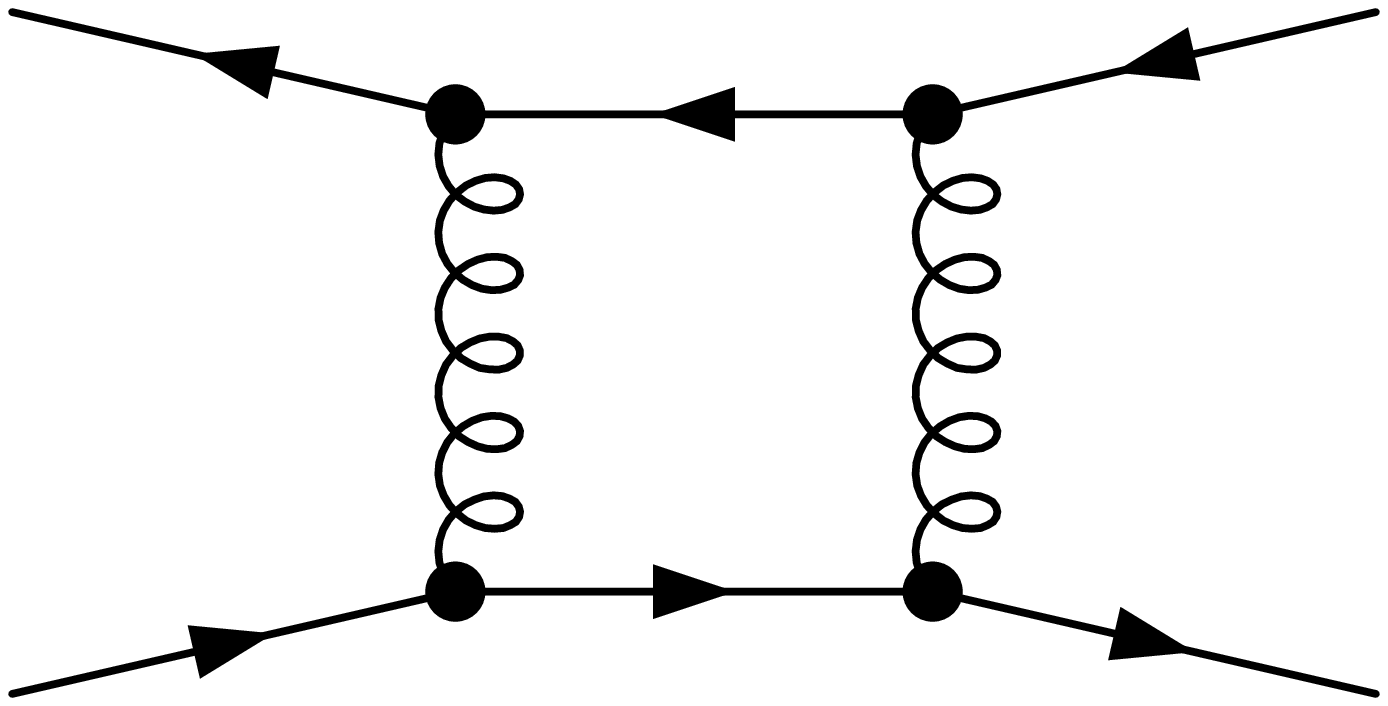}&
\hspace*{2mm}\includegraphics[width=2.4cm]{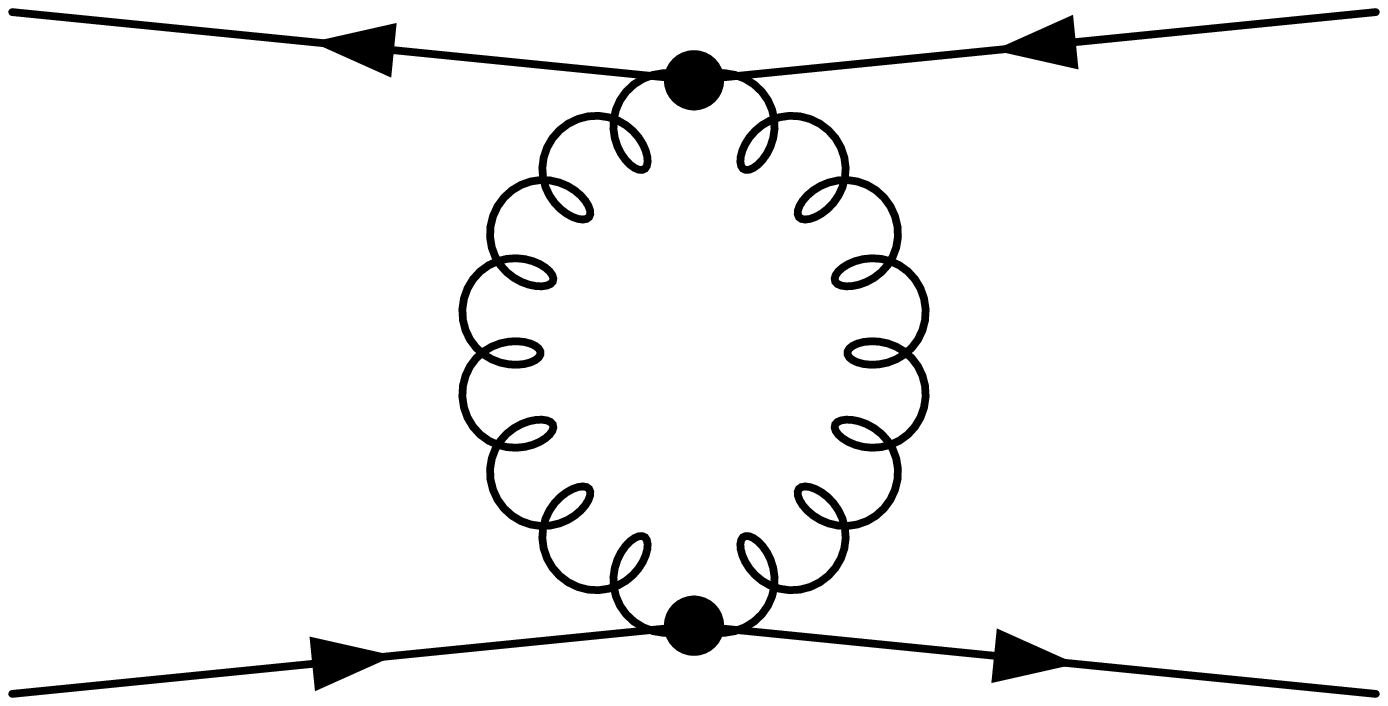}\\
(a)&\hspace*{2mm} (c) &\hspace*{2mm} (e)\\
\includegraphics[width=2.4cm]{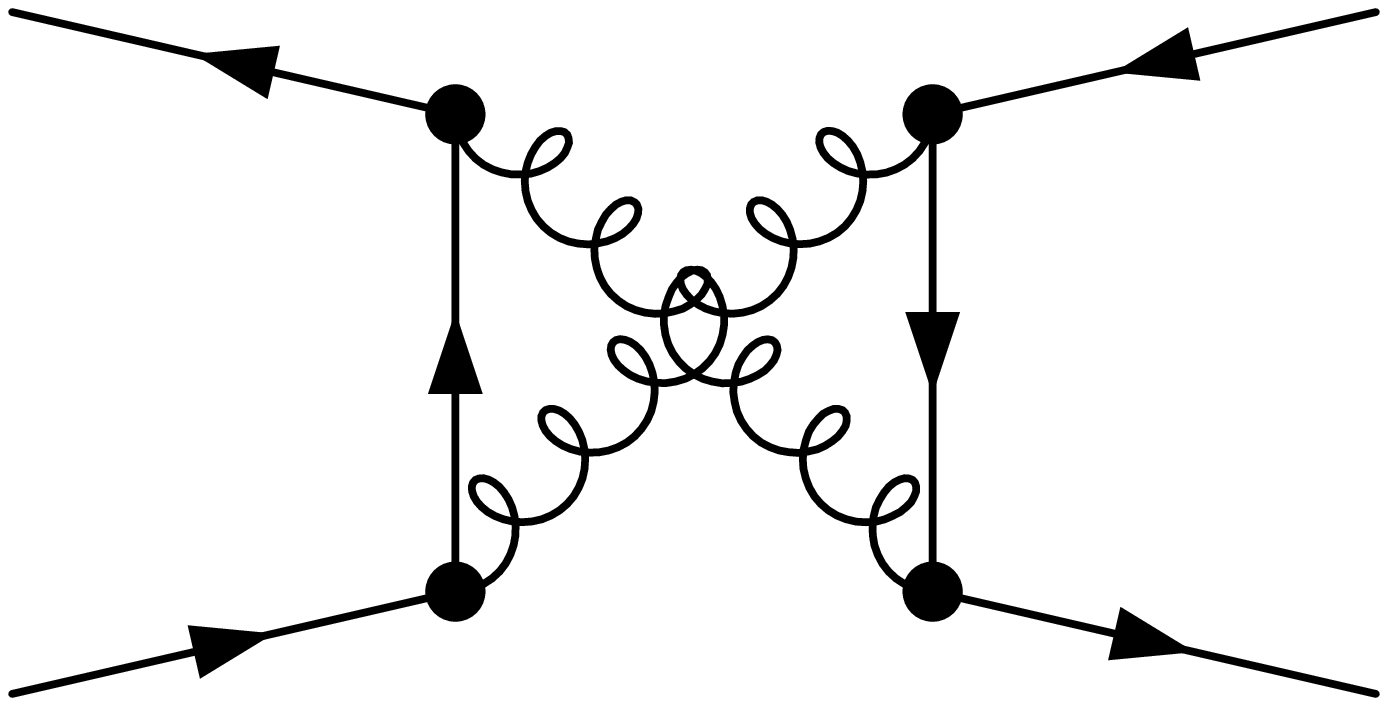}&
\hspace*{2mm}\includegraphics[width=2.4cm]{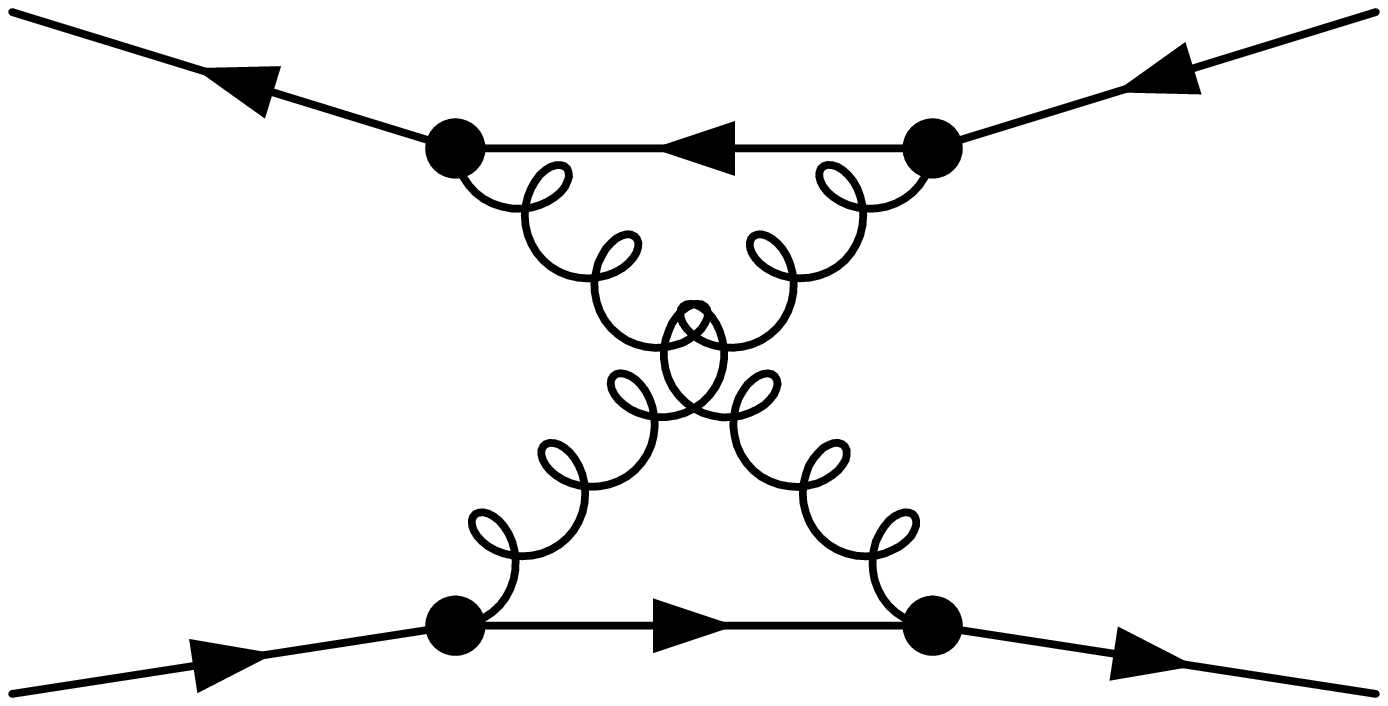}&
\hspace*{2mm}\includegraphics[width=2.4cm]{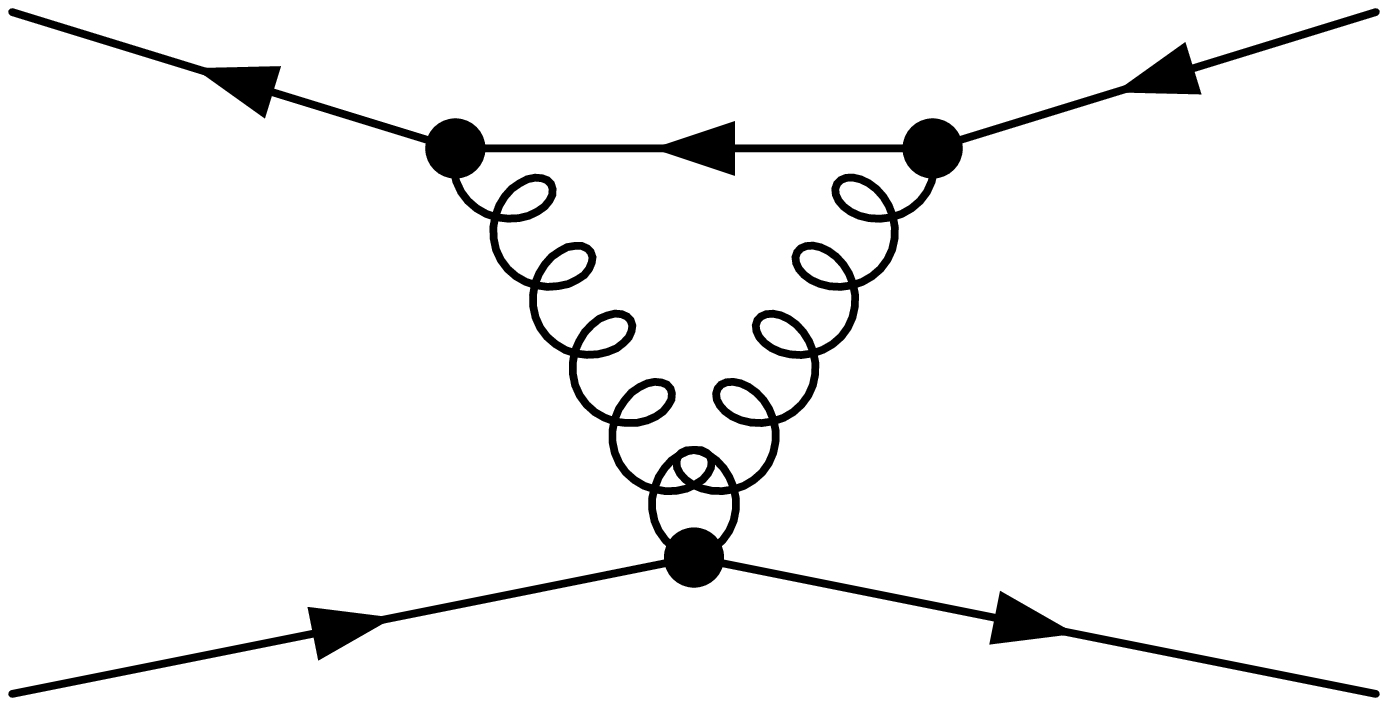}\\
(b)& \hspace*{2mm} (d) & \hspace*{2mm}(f)
\end{tabular}
\caption{\label{fig::fig1}  One-loop Feynman diagrams
contributing to the one-particle irreducible quark-antiquark
scattering amplitude in QCD [(a)-(d)] and NRQCD [(c)-(f)].}
\end{figure}

The coefficient $c_F$ parametrizes the quark anomalous chromomagnetic  moment.
It can be determined nonperturbatively by matching  the  lattice result for
particular splittings to the physical bottomonium  spectrum
\cite{Meinel:2010pv,Dowdall:2011wh}. The perturbative  evaluation of the
one-loop correction to $c_F$ \cite{Hammant:2011bt} is  in  good agreement with
the nonperturbative result. The Wilson coefficient of the effective four-quark
interaction however can only be obtained perturbatively. It vanishes in the Born
approximation and is determined by matching the one-particle irreducible
quark-antiquark  scattering amplitudes in QCD   and NRQCD; see Fig.~\ref{fig::fig1}.
The matching does not depend on the choice of soft kinematical variables and

becomes particulary simple when the amplitude is computed at the quark-antiquark
threshold and vanishing momentum transfer. In this case the one-loop
full QCD amplitude is
\begin{eqnarray}
 M_{\rm 1PI}^{\rm QCD} &=&
\frac{C_F\alpha_s^2}{m_q^2}\left[{C_A\over 2}\log\left({m_q\over\lambda}\right)
+\left(\ln 2-1\right)T_F\right.
\nonumber\\&+&\left.\left(1-\frac{2\pi m_q}{3\lambda}\right)C_F
      \right]\psi^\dagger {\bfm \sigma}\psi \chi_c^\dagger {\bfm \sigma}\chi_c,
\nonumber \\
\label{eq::ampqcd}
\end{eqnarray}
where  $C_A=N_c$, $T_F=1/2$, and we introduced a small auxiliary gluon mass
$\lambda$ to regulate the infrared divergence.  The power enhanced $1/\lambda$
term corresponds to the Coulomb singularity of the threshold amplitude, while the
term proportional to $T_F$ is due to the two-gluon annihilation of the
quark-antiquark pair.

\begin{table}[t]
  \begin{ruledtabular}
    \begin{tabular}{l|lc}
    \multicolumn{2}{c}{Experiment}  &\\
      \hline
      Babar, $\Upsilon(3S)$ decays\cite{Aubert:2008ba} & $71.4^{+2.3}_{-3.1}({\rm stat})\pm 2.7({\rm syst})$&\\
      Babar, $\Upsilon(2S)$ decays \cite{Aubert:2009as} & $66.1^{+4.9}_{-4.8}({\rm stat})\pm 2.0 ({\rm syst})$&\\
      Belle, $h_b(1P)$ decays \cite{Mizuk:2012pb} & $57.9\pm 2.3$&\\
      PDG average \cite{Beringer:1900zz} & $62.3\pm 3.2$&\\
      \hline\hline
      \multicolumn{2}{c}{Theory} & \\
      \hline
      NRQCD, NLL \cite{Kniehl:2003ap} &  $41\pm 11{\rm (th)}^{+9}_{-8}(\delta\alpha_s)$   & \\
      Lattice NRQCD ${\cal O}(v^4)$ \cite{Dowdall:2011wh}& $70\pm 9$ & \\
      Lattice NRQCD ${\cal O}(v^6)$ \cite{Dowdall:2013jqa}& $62.8\pm 6.7$  & \\
      Lattice QCD  \cite{Burch:2009az}& $54.0\pm 12.4^{+1.2}_{-0.0}$ & \\
      Lattice NRQCD, this work & $52.9\pm 5.5$   &\\
    \end{tabular}
    \end{ruledtabular}
    \caption{\label{tab::tab1} Results of high-precision experimental and
    theoretical determinations of the bottomonium hyperfine splitting in MeV.}
\end{table}

On the other hand the  lattice NRQCD result for the one-loop amplitude
can be written as follows
\begin{eqnarray}
M_{\rm 1PI}^{\rm NRQCD} &=&
\frac{C_F\alpha_s^2}{m_q^2}\left[-\left(\delta
+{1\over 2}\ln\left(a\lambda \right)\right)C_A
\right.
\nonumber \\
&-&\left.\frac{2\pi m_q}{3\lambda}C_F\right]\psi^\dagger {\bfm \sigma}\psi
\chi_c^\dagger {\bfm \sigma}\chi_c
+{\cal O}\left(a^2\right),
\label{eq::ampnrqcd}
\end{eqnarray}
where the nonlogarithmic non-Abelian term $\delta$ depends on a particular
realization of the lattice action. To match Eqs.~(\ref{eq::ampqcd}) and
(\ref{eq::ampnrqcd}) we add to the NRQCD Lagrangian the four-quark operator
with coefficient
\begin{equation}
d_\sigma=\alpha_s\left[\left(\delta +{1\over 2}L\right)C_A
+\left(\ln 2-1\right)T_F+C_F\right],
\label{eq::ds}
\end{equation}
where  $L=\ln(a m_q)$.
The main problem is  therefore in determination of the coefficient $\delta$. The
asymptotic expansion of the lattice loop integrals about the continuum limit
\cite{Becher:2002if} can in principle be used to get this coefficient in a
closed analytic form. Since the heavy quark mass is not a dynamical scale in
NRQCD, the parameter of the expansion in our case is  $a\lambda$. The idea of
the method is to split the integration over the virtual momentum $l$ into the
contributions of the hard region with $l\sim 1/a$ and the soft region with
$l\sim \lambda$. In the hard region the integrand is expanded in $a\lambda$ and
$\lambda/l$ and reduces to the lattice tadpole integrals.  In the soft region
the integrand is expanded in $al$ and $a\lambda$ and reduces to the continuum
NRQCD Feynman integrals. As a result of the scale separation  the hard (soft)
contribution in general has  spurious  infrared (ultraviolet) logarithmic
divergences and has to be regulated. In the total result for a given lattice
loop integral the dependence on the regulator cancels out leaving the asymptotic
series in   $a\lambda$ which includes the logarithmic terms [{\it cf.}
Eq.~(\ref{eq::ampnrqcd})]. We emphasize that the expansion about
the continuum limit is a formal tool to get the NRQCD loop integrals as series in
$a$ and facilitate the matching procedure, while the lattice NRQCD is a valid
nonrelativistic effective theory only for $a\gg 1/m_q$. Note that
Eq.~(\ref{eq::ampnrqcd}) has only a logarithmic singularity in $a$. In higher
orders of the NRQCD expansion in $1/m_q$ the amplitude includes also the terms
with a negative power of $a$. Such  $1/(am_q)^n$ terms are more singular in the
formal continuum limit but are power suppressed with respect to
Eq.~(\ref{eq::ampnrqcd}) in the  region where lattice NRQCD is applied.

Let us consider first  a  ``naive''   lattice action  with
no improvement for gluonic and heavy  quark fields (see, {\it e.g.}
\cite{Rothe:kp,Eichten:1989kb}). The gluonic field  tensor of the  NRQCD
chromomagnetic interaction in the naive action is expressed   through the
commutator of the left-right symmetrized covariant lattice derivatives. In this
case we obtain
\begin{equation}
\delta^{\rm naive}=-{7\over 3}+28\pi^2 b_2-256\pi^2 b_3=0.288972\ldots,
\label{eq::delnaive}
\end{equation}
where the irrational constants $b_2 = 0.02401318\ldots$, $b_3 =
0.00158857\ldots$ parametrize the lattice tadpole integrals and can be computed
with arbitrary precision \cite{Becher:2002if}.
We however need the above coefficient  for the improved lattice action which is
used in real simulations. Analytic  calculation with an improved action is not
optimal since the Feynman rules in this case  become extremely cumbersome. We
bypass this problem by using a semianalytic approach. Indeed the difference
between the Wilson coefficients for the improved and naive lattice actions
$\Delta \delta$ remains finite in the limit $\lambda\to 0$ and can be directly
obtained by numerical evaluation of the corresponding lattice loop integrals
with sufficiently small $\lambda$ (a finite infrared regulator is necessary for
the stability of numerical  integration). For the numerical implementation of
the improved lattice action  Feynman rules we use HiPPy/HPsrc code
\cite{Hart:2009nr}. However in contrast to the standard implementation the color
space reduction is performed analytically with the help of the program
COLOR~\cite{vanRitbergen:1998pn} before the numerical integration is done by the
CUBA integrator library~\cite{Hahn:2004fe}. This  greatly reduces the runtime
and allows for a separate treatment of the contributions of independent color
group structures which have different infrared properties, {\it cf.}
Eq.~(\ref{eq::ampnrqcd}). The whole process of the calculation is fully
automated. In the case of the HPQCD action \cite{Dowdall:2011wh} we get $\Delta
\delta= -0.1444(28)$ corresponding to
\begin{equation}
\delta=0.1446(28)\,.
\label{eq::delHPQCD}
\end{equation}

\begin{figure}
\includegraphics[width=8cm]{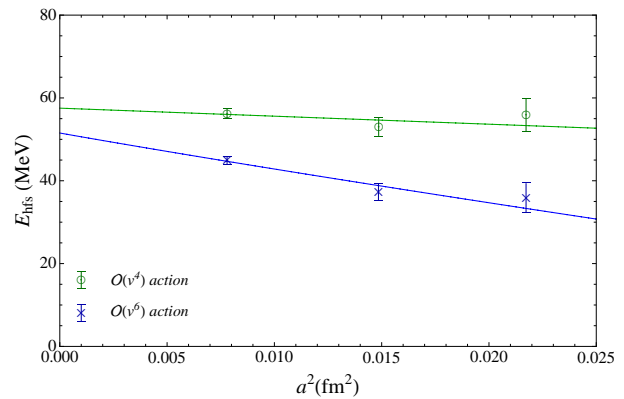}
\caption{\label{fig::fig2} The result of the lattice simulation of the
bottomonium hyperfine splitting with ${\cal O}(v^4)$  action \cite{Dowdall:2011wh}
and  ${\cal O}(v^6)$  action \cite{Dowdall:2013jqa}. The error bars are
explained in the text. The solid lines correspond to the central values of the
constrained fit.}
\end{figure}

\noindent
Note that since $d_\sigma=0$ in the Born approximation, we have to
perform neither the strong coupling constant renormalization nor the lattice tadpole
improvement. We made a few cross-checks of the calculation. For the naive action
the numerical evaluation agrees with the gauge-invariant analytic result of the
asymptotic expansion for small values of $\lambda$. The logarithmic part of
$d_\sigma$ is in agreement with the renormalization group analysis. The
nonrelativistic renormalization group predicts the all-order dependence of the
Wilson coefficients on  $\mu_f$ \cite{Penin:2004ay,Penin:2004xi}. In the leading
logarithmic approximation they read
\begin{equation}
d_\sigma^{LL}={C_A\over \beta_0-2C_A} \left(z^{-2C_A}-z^{-\beta_0}\right),
\quad c_F^{LL}=z^{-C_A},
\label{eq::LL}
\end{equation}
where $\beta_0=11C_A/3-4T_Fn_l/3$ is the one-loop QCD $\beta$ function, $n_l=4$
is the number of light flavors, and
$z=(\alpha_s(\mu_f)/\alpha_s(m_q))^{1/\beta_0}$. In lattice NRQCD the
factorization scale should be identified with inverse lattice spacing $\mu_f\sim
1/a$. By reexpanding the leading logarithmic result we obtain
\begin{eqnarray}
{d_\sigma^{LL}\over\pi}&=&
{\alpha_s\over \pi}{C_A\over 2}L-
\left({\alpha_s\over \pi}\right)^2{\left(\beta_0+C_A\right)C_A\over 4}L^2+\ldots,
\nonumber \\
c_F^{LL}&=&1
-{\alpha_s\over \pi}{C_A\over 2}L+
\left({\alpha_s\over \pi}\right)^2{\left(\beta_0+C_A\right)C_A\over 8}L^2+\ldots,
\nonumber \\
&&
\label{eq::LLexp}
\end{eqnarray}
in agreement with Eq.~(\ref{eq::ds}).

Let us now compare our result with the previous calculation
\cite{Hammant:2011bt}. In this paper a different basis of the four-quark
operators is used and the  Wilson coefficient $d_\sigma/\alpha_s$ should be
identified with the linear combination  ${9\over 8}(d_1-d_2)$ (see
\cite{Hammant:2013sca} for the consistent analytical expressions). We find that
the nonannihilation constant term of the QCD amplitude in \cite{Hammant:2011bt}
is smaller than the one in  Eq.~(\ref{eq::ampqcd}) by a factor of $3$. The comparison
of the NRQCD part of the result is not straightforward since in
\cite{Hammant:2011bt,Hammant:2013sca} it has been evaluated numerically for
three different lattice spacings keeping the full dependence on $m_q$. This
dependence includes power suppressed terms as well as the  linear term  from the
lattice cutoff of the Coulomb pinch contribution. For the lattice spacing used
in  real simulations $a\sim 1/(v m_q)$ the power suppressed terms  are of the
same magnitude as the generic one-loop relativistic corrections and are beyond
the accuracy of our analysis.  On the contrary the lattice artifacts are
numerically significant. The linear term associated with the Coulomb pinch can
be estimated by cutting the corresponding continuum NRQCD one-loop integral at
the scale $\pi/a$.  This gives an additional  contribution to  Eq.~(\ref{eq::ds}),
\begin{equation}
-\nu{8\over 3}{C_F\alpha_s\over\pi}am_q \approx -0.94\,\alpha_sam_q\,,
\label{eq::art}
\end{equation}
where the factor $\nu=0.831\ldots$ adjusts the analytical result for the
integral over spherical momentum domain to the integral over the Brillouin zone.
The  numerical result  of Ref.~\cite{Hammant:2011bt} suggests a significantly 
larger negative coefficient of about  $-1.8$. Moreover in the threshold region 
the multiple Coulomb gluon exchange contributions  are not parametrically 
suppressed and the modification of the Coulomb bound state dynamics on the finite 
lattice is not accounted for by the one-loop analysis. It may change the numerical 
coefficient of the linear term  and generates all-order contributions in 
$\alpha_s am_q$. This means that (i) the one-loop matching does not remove the
linear lattice artifact at ${\cal O}(\alpha_s)$ and (ii) one cannot use the
finite lattice spacing  $a\sim 1/(\alpha_s m_q)$ as a Wilsonian cutoff for
NRQCD as it was done in \cite{Hammant:2011bt,Hammant:2013sca}.  Thus all the
lattice  artifacts  should be removed nonperturbatively by numerical
extrapolation of the lattice data to $a=0$ \cite{Dowdall:2011wh,Dowdall:2013jqa}.

\begin{table}[t]
  \begin{ruledtabular}
    \begin{tabular}{l|cc|cc}
      & ${\cal O}(v^4)$ action && ${\cal O}(v^6)$ action &\\
     \hline
      Discretization error &   2.6 && 3.1 &\\
      Relativistic corrections &  6.0 && 1.8 &\\
      Radiative corrections &  4.8 && 4.3 &\\
      \hline
      $E_{\rm hfs}$ & $57.5$ && $51.5$ &\\
    \end{tabular}
    \end{ruledtabular}
    \caption{\label{tab::tab2} The central value and the error budget  for the
    lattice NRQCD determination of the bottomonium hyperfine splitting with
    ${\cal O}(v^4)$  action \cite{Dowdall:2011wh} and  ${\cal O}(v^6)$ lattice
    action \cite{Dowdall:2013jqa} in MeV.}
\end{table}

Now we are in a position to apply our result to the analysis of the hyperfine
splitting. The  contribution of  the four-quark interaction to $E_{\rm hfs}$
reads
\begin{equation}
\Delta E_{\rm hfs}=-d_\sigma{4 C_F\alpha_s\over m_q^2}|\psi(0)|^2,
\label{eq::DelE}
\end{equation}
where $\psi(0)$ is the wave function of the  quarkonium ground state at the
origin. Equation~(\ref{eq::DelE}) should be added to the result of the lattice
simulation with the one-loop Wilson coefficient  $c_F$ and no four-quark
contribution included. Such a result is available for  the ${\cal O}(v^4)$
action \cite{Dowdall:2011wh} and for the ${\cal O}(v^6)$ action
\cite{Dowdall:2013jqa}.  For the numerical analysis of Eq.~(\ref{eq::DelE}) we
use the nonperturbative lattice result for $\psi(0)$ \cite{Dowdall:2011wh}. To
make our analysis self consistent we adopt the value of the bottom quark mass
$m_b$ and the value of  the strong coupling constant $\alpha_V$ renormalized  in
the static potential scheme at the scale $\pi/a$ from
Ref.~\cite{Dowdall:2011wh}. The numerical result for the hyperfine splitting is
presented in Fig. \ref{fig::fig2} as a function of $a^2$ for  three different
lattice spacings and two different lattice actions. The error bars of each point
include the statistical error and the uncertainty in the value of the lattice
spacing from \cite{Dowdall:2011wh,Dowdall:2013jqa} as well as the high-order
$a$-dependent radiative corrections that are estimated by the size of the
double-logarithmic two-loop terms in Eq.~(\ref{eq::LLexp}).
The use of relatively  large values of the  lattice spacing $a\sim 1/(vm_b)$
ensures the suppression of the unphysical $1/(am_b)^n$ contributions, which
become important at $a\sim 1/m_b$ \cite{Dowdall:2011wh,Hammant:2013sca}. At the
same time it results in sizable lattice artifacts, which cannot be removed by
finite order matching due to the  all-order character of the Coulomb binding
effects. To minimize this effect the result is  numerically extrapolated to
$a=0$ \cite{Dowdall:2011wh,Dowdall:2013jqa}. The extrapolation below  $a\sim
1/m_b$ in this case is justified since the numerical effect of the $1/(am_b)^n$
terms on the data points is small. To perform the extrapolation we use a
constrained fit of the data points \cite{Lepage:2001ym}  by a polynomial in $a$
with  vanishing linear term.   The inclusion  of the
linear and $1/(am_b)^n$  terms in the fit is discussed below.
To estimate the coefficients of the higher order
terms in the lattice spacing we represent the result  of the fit as $1+(\Lambda
a)^2+{\cal O}(a^3)$, where $\Lambda$ is the mass  scale characterizing  the
approach of the lattice approximation to the continuum limit. The priors for the
coefficients of the $a^n$ terms with $n>2$ in the constrained fit  are then
given by the intervals $\pm\Lambda^n$.  Numerically we get $\Lambda\approx
360$~MeV for the ${\cal O}(v^4)$ and $\Lambda\approx 790$~MeV for the ${\cal
O}(v^6)$ case. Because of a slower approach to the continuum limit the extrapolation
 error for ${\cal O}(v^6)$ action turns out to be larger. This may be related to
the fact that the contribution of the operators of higher dimension is more
sensitive to the ultraviolet momentum region. Therefore the currently unknown
${\cal O}(\alpha_s v^6)$  matching corrections in this approximation can be
substantial. We checked that the inclusion of the $1/a^n$  terms with the priors
${\alpha_s\over\pi} ({\pi\over m_b})^n$ into the constrained fit changes the
result within the  extrapolation error intervals.

In general the Coulomb binding effects give rise to a linear dependence of the
lattice data on $a$ which can be roughly estimated by the one-loop
result~(\ref{eq::art}). A more refined estimate  can be obtained by including the 
linear term $c_{l}\alpha_s am_q$  into the fit of the lattice data. For  the prior
$|c_{l}|<1$ the constrained fit gives $c_l\approx -0.25$ for both actions, 
which is  two times smaller than the one-loop estimate $c_{l}\approx -0.5$
corresponding to Eq.~(\ref{eq::art}). At the same time the
extracted  value of the hyperfine splitting is increased within the extrapolation 
error interval by approximately $2.5$~MeV.

The total error budget of
our estimate is given in Table~\ref{tab::tab2}. Besides the discretization errors
discussed above it includes the uncertainty due to high-order relativistic and
radiative corrections. For a conservative estimate of the radiative corrections
we take the value of the double-logarithmic two-loop terms at the soft
factorization scale $\mu_f\approx \alpha_s m_b$ dictated by the bound state
dynamics. In Table~\ref{tab::tab2} this uncertainty is combined with the
numerical error in the one-loop coefficient $c_F$ \cite{Dowdall:2013jqa}.  Our
estimate of the relativistic corrections for the ${\cal O}(v^4)$ action is based
on the difference between the  ${\cal O}(v^4)$ and ${\cal O}(v^6)$ results in
the continuum limit. For the  ${\cal O}(v^6)$ action we multiply this
uncertainty by $\alpha_s$  evaluated at the soft renormalization scale to take
into account the previously discussed missing matching corrections. The larger
discretization uncertainty  balances the smaller relativistic corrections in the
 ${\cal O}(v^6)$  case and both actions provide comparable total errors. Since
the structure of the relativistic corrections and the behavior of the results at
finite lattice spacing are significantly different for the two actions, we
consider the corresponding uncertainties as uncorrelated and  take the weighted
average of the results as the best estimate. At the same time the uncertainty
due to the high-order purely radiative corrections is treated as correlated
between the two actions. Our final result for the hyperfine splitting reads
\begin{equation}
E_{\rm hfs}=52.9\pm 5.5~{\rm MeV}.
\label{eq::fin}
\end{equation}
We now can compare our estimates to the available theoretical and
experimental results in Table~\ref{tab::tab1}. Our result for both
${\cal O}(v^4)$ and ${\cal O}(v^6)$ actions (Table~\ref{tab::tab2}) are below
the ones of the previous lattice NRQCD analysis
\cite{Dowdall:2011wh,Dowdall:2013jqa} by approximately $12$~MeV. About $5$~MeV
of the difference is due to the error in the one-loop QCD amplitude calculation
\cite{Hammant:2011bt}. The remaining discrepancy is related to the different
procedure of extrapolation to $a=0$. The analysis~\cite{Dowdall:2011wh,Dowdall:2013jqa}
implies that the one-loop matching \cite{Hammant:2011bt} removes the linear
artifact form the lattice data. However, as it was pointed out above,
the  one-loop calculation can only be used for a rough estimate of the linear term
due to the  all-order character of the Coulomb binding effects.
We therefore determine the corresponding coefficient by a constrained fit of
the lattice data with the prior set by the one-loop result. Moreover the numerical result~\cite{Hammant:2011bt} suggests a significantly larger  value of the linear
term than what follows from our analytic calculation and from the fit of the lattice data,
which leads to a sizable difference of the extrapolation results.

With the new value of the four-quark Wilson coefficient the lattice NRQCD 
prediction~(\ref{eq::fin}) agrees within the error bars with the next-to-leading 
logarithmic (NLL) perturbative QCD result
\cite{Kniehl:2003ap}. Its  central value  practically coincides with that of
the full lattice QCD simulation \cite{Burch:2009az},  though  the uncertainty of
the latter is significantly larger. This may indicate that the matching
of the lattice NRQCD to full QCD is now done properly. On the experimental side
our result  strongly favors the value obtained by the Belle collaboration, which has
the lowest reported uncertainty. Thus we have reconciled the theoretical
predictions of the lattice and continuum QCD as well as the most accurate
experimental data.

\vspace{5mm}

\begin{acknowledgements}
The work  was supported in part by NSERC.
The work of A.P. is supported by the Perimeter Institute for Theoretical Physics.
Research at the Perimeter Institute is supported by the Government of Canada
through Industry Canada and by the Province of Ontario through the Ministry
of Research and Innovation.
\end{acknowledgements}


\end{document}